\documentclass[10pt,twocolumn,prd, showpacs,amssymb,preprintnumbers,nofootinbib,
superscriptaddress]{revtex4}

\usepackage{amsmath}
\usepackage{graphicx}
\usepackage{latexsym}
\usepackage{amsfonts}
\usepackage{url,hyperref}
\usepackage{bm}
\usepackage{bbm}

\usepackage{fancyhdr}

\begin{document}

\begin{figure}[t]
\vspace{1.4cm} \hspace{16.15cm}
\end{figure}

\newcommand{\vp}{\varphi}
\newcommand{\nn}{\nonumber\\}
\newcommand{\beq}{\begin{equation}}
\newcommand{\eeq}{\end{equation}}
\newcommand{\bed}{\begin{displaymath}}
\newcommand{\eed}{\end{displaymath}}
\def\bea{\begin{eqnarray}}
\def\eea{\end{eqnarray}}
\newcommand{\veps}{\varepsilon}
\def\slash{\not\!}

\title{Lepton number violation via intermediate black hole processes}
\author{Jason Doukas}
  \email[Email: ]{j.doukas@physics.unimelb.edu.au}
   \affiliation{School of Physics, Research Centre for High Energy Physics,\\
    University of Melbourne, Parkville, Victoria 3010, Australia.}
\author{S. Rai Choudhury}
  \email[Email: ]{src@physics.du.ac.in}
    \affiliation{Department of Physics \& Astrophysics,\\
     University of Delhi, Delhi 110007, India,}
\author{G. C. Joshi}
  \email[Email: ]{joshi@tauon.ph.unimelb.edu.au}
   \affiliation{School of Physics, Research Centre for High Energy Physics,\\
    University of Melbourne, Parkville, Victoria 3010, Australia.}
\begin{abstract}
Black holes at the TeV scale are investigated in the extra large
dimension scenario. We interpret the lightest black hole
excitation as a singlet scalar field, and show how interaction
terms can be appended to the standard model at the dimension five
non-renormalizable level. Lepton family number violation is
natural in this model. Muon magnetic moment, and neutrino masses
are investigated. We also present a quantization scheme in n
dimensions.
\end{abstract}

\date{\today} \maketitle

\section{Introduction}
The existence of black holes at the M$_P \sim$ TeV range has been
conjectured \cite{Argyres:1998qn,
Dimopoulos:2001hw,Giddings:2001bu} within recently proposed
scenarios of extra large dimensions \cite{Arkani-Hamed:1998rs,
Arkani-Hamed:1998nn, Antoniadis:1998ig}. In light of planned
searches at the LHC \cite{Tanaka:2004xb}, an effective field
theory for black hole processes has been designed
\cite{Bilke:2002rf, Choudhury:2003xf}. This approach is superior
to other avenues for black hole analysis in the respect that it
allows one to calculate interference effects of black
hole processes with relative ease.\\
In this paper we show how black hole excitations can be
incorporated into the standard model at the dimension five
non-renormalizable level. We calculate the corrections that the
muon magnetic moment receives from a black hole at the TeV
scale and show that black hole corrections do not spoil the smallness of the neutrino masses. \\
It is assumed that their are n additional compact spatial
dimensions possibly at the size of a millimeter and that the
fundamental Planck scale is comparable to the electroweak scale
\cite{Arkani-Hamed:1998rs}. Black hole solutions on such higher
dimensional backgrounds are rare. However, in the case when $\rm
R_{s}\ll R$ where $\rm R_{s}$ is the Schwarzschild radius and R
the characteristic size of the extra dimensions, we can
approximate the dimensions as asymptotically flat and then use the
results of Myers and Perry \cite{Myers:1986un}. To wit $\rm
R_{s}\sim \frac{1}{M_{ew}}
\left(\frac{M_{bh}}{M_{ew}}\right)^{1/(n+1)}$, and $\rm R\sim
10^{30/n} ~TeV^{-1}$, so we would need a black hole of mass $\rm
\gtrsim 10^{30} ~TeV$ before this assumption breaks down.
Typically, we will be dealing with black holes of $\rm M_{bh}\sim 1~TeV$.\\
The black hole in this paper is treated as a particle quantized in
mass \cite{Bekenstein:1974jk}. Originally this quantization was
constructed in 4 dimensions. In Appendix 1
we present the analogous result for a charged spin zero black hole
in n+3 spatial dimensions.\\
\section{Summary of the model} 
\begin{figure}[h]
\centering
\includegraphics{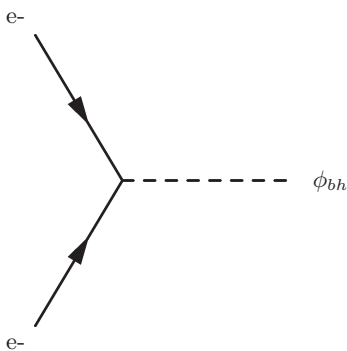}
\caption{Black hole production.} \label{fig:production}
\end{figure}

In essence the philosophy of
the model is to interpret each quantized black hole excitation as
an independent quantum field. The lightest such excitation has
zero charge and zero angular momentum and thus corresponds to a
neutral scalar field. The original ansatz \cite{Bilke:2002rf} was
constructed for the interaction between two charged fermions and a
doubly charged black hole. It consisted of a Yukawa Lagrangian of
the type:
\begin{eqnarray}\label{eqn:Lagrangian2}
&&{\cal L}_{\rm int} = i\rm k_{\rm eff} M_{\rm bh} \phi_{\rm bh}
\overline \Psi_f \hat C \Psi_f + h.c.\,,
\end{eqnarray}
where $\rm \hat C=i \gamma_2 \mathcal{K}$ is the charge
conjugation operator, $\mathcal{K}$ being complex conjugation and
$\Psi_f$ is the fermion field.
Similarly, one can write an effective Lagrangian for a neutral
scalar black hole by removing the $\rm \hat C$ operator:
\begin{equation}\label{eqn:LagrangianScalar} \mathcal{L}_{\rm int}=
i\rm k_{\rm eff}M_{\rm bh}\phi_{\rm bh}\overline{\Psi}_{f}\Psi_{f}
+h.c.
\end{equation}
Photon-black hole interactions have also previously been
considered \cite{Choudhury:2003xf}:
\begin{equation}
\mathcal{L} _{int} = \rm \frac{k_{eff}}{M_{p}}  \phi_{bh}
F_{\mu\nu} F^{\mu\nu} .
\end{equation}
In this paper we present a rationale for the choice of these black
hole interactions. That is, they all arise from non-renormalizable
dimension five operators that respect the standard model
symmetries. Thus one imagines that this effective theory is the
remnant of some quantum gravity theory broken at the $\rm M_p$ scale.\\
For simplicity we focus on the phenomenology of a single neutral
scalar black hole, $\rm \phi_{bh}$. This is a singlet under the
standard model gauge group. Therefore at the dimension five level
we have:
\begin{equation}\label{eqn:LagrangianDim5}
\mathcal{L}=\mathcal{L}_{SM}+\frac{g'}{\Lambda}\mathcal{L}_{SM}
\phi_{bh}+(D\phi_{bh})^2-V(\phi_{bh})+h.c.,
\end{equation}
where $\rm g'\mathcal{L}_{SM} \phi_{bh}$ is symbolic for all
standard model operators times a $\rm\phi_{bh}$ operator with
possibly different coupling constants $g'$ on each term and D is
the covariant derivative. We expect that the theory will be cut
off at the fundamental planck scale $\rm
\Lambda=M_{p}$, where a new theory that accommodates gravity will take over.\\
It is the second term of (\ref{eqn:LagrangianDim5}) that contains
the black hole interactions used in previous work. For example one
sees that the interaction in equation (\ref{eqn:LagrangianScalar})
can be obtained from the $\rm
\frac{g'}{M_p}\overline{L}\Phi_{H}e_R\phi_{bh}$ term after Higgs
breaking $<\phi_H>=v$,
\begin{eqnarray*}
\rm
\frac{g'}{M_p}\overline{L}\Phi_{H}e_R\phi_{bh}+h.c.&\rightarrow&\rm
\rm k_{eff}M_{bh}\overline{\Psi}_e\Psi_e \phi_{bh},
\end{eqnarray*}
if we make the identification $\rm k_{eff}=\frac{g'
\emph{v}}{M_pM_{bh}}$.\\
The authors \cite{Bilke:2002rf} have determined $\rm k_{\rm eff}$
by comparing the black hole effective production cross section
with the geometrical cross section $\rm\sigma_{geom}=\pi R_{bh}^2$
on resonance, ie multiplied by a dimensionless generalized
function peaked at the com energy $\rm \sqrt{s}=M_{bh}$:
\begin{equation}
\rm\sigma_{geom}=\pi R_{bh}^2 M_{bh}\delta(\sqrt{s}-M_{bh}),
\end{equation}
The relevant effective production process is shown in FIG.
\ref{fig:production}.
\newline
This has the cross section,
\begin{equation}
\rm \sigma_{eff} =\frac{1}{4} |k_{eff}|^2
M_{bh}\delta(\sqrt{s}-M_{bh}).
\end{equation}
After equating $\rm \sigma_{eff}$ and $\rm \sigma_{geom}$ the
relationship $\rm k_{eff}=2R_{s}$ is made.
The radius of the n+3 dimensional Schwarzschild black hole is
solved in \cite{Myers:1986un}, thus,
\begin{eqnarray}\label{eqn:2}
\rm k_{eff}=\rm \frac{2}{\sqrt{\pi}
M_{p}}\left[\frac{M_{bh}}{M_{p}}\frac{8\Gamma(\frac{n+3}{2})}{n+2}\right]^{1/(n+1)}.
\end{eqnarray}
This gives the same $\rm k_{eff}\propto \frac{1}{M_p}$ dependence
that we expect if the theory is cut off at the $\rm \Lambda=M_p$
scale. The two approaches are consistent if the black hole mass is
at the electroweak scale. From here on we assume that $\rm
k_{eff}=g'/M_p$. Where $\rm g'$ is a coupling constant dependent
on the
particular interaction under investigation.\\
%
\section{Lepton number violation}\label{sec:LeptonViolation}
\begin{figure}[h]

\includegraphics{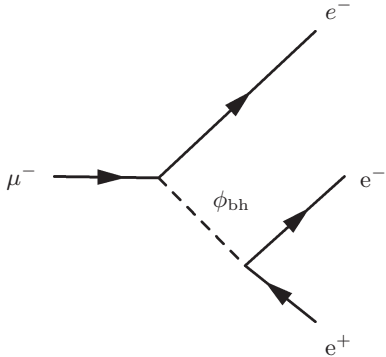}

\caption{Feynman diagram for the muon decay via the neutral scalar
black hole interaction.} \label{fig:Muondecay}
\end{figure}

It has been conjectured that black hole processes will violate
certain approximate global symmetries like lepton number
\cite{Dimopoulos:1979ma,Witten:2000dt,Zichichi:1977ri}. Presently,
we wish to extend the original work
\cite{Bilke:2002rf,Choudhury:2003xf} to include lepton family
number violation. This is implemented in equation
(\ref{eqn:LagrangianDim5}) by allowing the usual 3 generations of
fermion fields in $\mathcal{L}_{SM}$.
Consider the terms
\begin{eqnarray}
\mathcal{L}=\rm\ldots+\frac{1}{M_p} \overline{L}_i \Phi_H
g'_{ij}e_{Rj}\phi_{bh}+\overline{L_i}\Phi_H\lambda_{ij}e_{Rj},
\end{eqnarray}
where a summation over generations is understood ($i,j \in 1,2,3
$). Now after Higgs breaking and rotating the weak eigenstates
into mass eigenstates, ie $\rm \mathbbm{m}=\mathbbm{U}\lambda
\mathbbm{V}^{\dagger}$, this Lagrangian becomes
\begin{eqnarray}
\rm \mathcal{L}\rightarrow\overline{e}_{Li}
k_{ij}e_{Rj}\phi_{bh}+m_i\overline{e}_{Li}e_{Ri}.
\end{eqnarray}
Importantly $\rm \mathbbm{k}\equiv
\frac{\emph{v}}{M_p}\mathbbm{U}\mathbbm{g}'\mathbbm{V}^{\dagger}$
is not in general diagonal. Thus lepton family violating
interactions like $\rm k_{e\mu}\overline{\mu}e\phi_{bh}$ arise
naturally in this picture as off diagonal terms in $\rm
\overline{e}_{Li} k_{ij}e_{Rj}\phi_{bh}$. \\
\newline
We now consider the decay mode $\rm \mu^{-}\rightarrow
e^{-}e^{+}e^{-}$ \cite{PDBook}, which proceeds through the
processes shown in FIG. \ref{fig:Muondecay}. It has been said
\cite{Bilke:2002rf} that the experimental bound on this process
would require a Planck scale at the 100 TeV range. In our approach
we have an extra parameter in $\rm g'$, recall $\rm k=
\frac{g'}{M_p}$, if we want a TeV scaled Planck mass then we must
also tolerate $\rm g'\sim 10^{-3}-10^{-4}$ sized couplings.

\newpage
The decay rate is:
\begin{equation}\label{eqn:Gamma}
\rm \Gamma(\mu^{-}\rightarrow
e^{-}e^{+}e^{-})= m_{\mu}^5 k_{ee}^2k_{\mu e}^2,\\
\end{equation}
which puts a bound on the product of the two couplings,
\begin{equation*}\nonumber
\rm k_{ee}k_{\mu e}<1.5\times10^{-7}~TeV^{-2}.
\end{equation*}
\newline
\newline\newline
\newline\newline\newline
Since we take $\rm M_p\sim 1 ~TeV$ this means
\begin{equation}\label{eqn:coupling}
\rm g_{ee}g_{\mu e}<1.5\times10^{-7},
\end{equation}
which is consistent with a TeV scaled Planck mass. It is also
interesting that with tolerable tuning we could have one g rather
large $\sim 10^{-1}$ and the other small $\sim 10^{-6}$. Our model
would therefore be able to accommodate black holes that favor
certain processes. This does not contradict thermal arguments as
our understanding of Hawking radiation breaks down at the Planck
scale.\\
~\\ ~\\
\newline
\newline
\begin{widetext}
\section{Muon Magnetic moment correction}\label{sec:muon}
In this section we calculate the correction to the muon magnetic
moment from the diagram shown in FIG. \ref{fig:muon}.\\
\begin{figure}[h]
\centering
\includegraphics{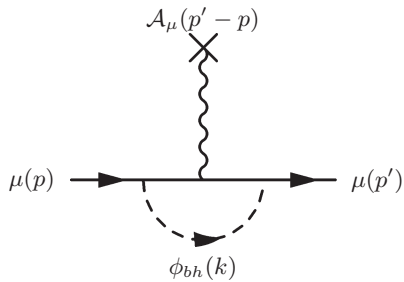}
 \caption{Muon magnetic moment correction}
\label{fig:muon}
\end{figure}
The matrix element for this process is:
\begin{equation}
\rm
\mathcal{M}=ek_{el}^{2}M_{bh}^{2}\overline{U}(p')\Lambda^{\mu}U(p)\mathcal{A}_{\mu}(p'-p),
\end{equation}
where,
\begin{equation} \rm \Lambda^{\mu}=
\frac{-i}{(2\pi)^{4}}\int\frac{dk^4}{k^{2}-M_{bh}^{2}+i\epsilon}\frac{
\slash{p'}-\slash{k}
+m_{l}}{(p'-k)^2-m_{l}^2+i\epsilon}\gamma^{\mu} \frac{
\slash{p}-\slash{k} +m_{l}}{(p-k)^2-m_{l}^2+i\epsilon},
\end{equation}
\end{widetext}and we have allowed for lepton number violating vertices, ie the internal fermion lines
in FIG. \ref{fig:muon} could be non-muonic charged leptons of mass $\rm m_l$.\\
This gives a corresponding correction to the muon magnetic moment
of
\begin{eqnarray*} \rm
a_{bh}&=&
\rm\frac{k_{el}^2m_{\mu}^2}{8\pi^2}\int^1_0dz\frac{z^2(\beta-z)}{(1-z)+\alpha
z^2},
\end{eqnarray*}
where $\rm\beta =1+\frac{m_l}{m_{\mu}}$ and $\rm\alpha=(\frac{m_l}{M_{bh}})^2$.\\
If one assumes a $\rm g_{\mu\mu}\lesssim 1$ the correction is of
order $\rm a_{bh} \sim 10^{-10}$, which is close to the current
level of deviation between standard model and experiment $\rm
|a_{exp}-a_{theory}|~ < ~42.6 \times 10^{-10}$
\cite{Czarnecki:2001pv}. Choosing $\rm g_{\mu\mu}\lesssim 1$ is
not inconsistent with the muon decay result (\ref{eqn:coupling}),
however it does require some tuning. Assuming that all couplings
are of the same order brings us to corrections of the size
$10^{-15}$. In any case it is not possible for perturbative $\rm
g$ to produce corrections that would contradict experimental
results. Specifically, an overestimate can be obtained by taking
$\rm g=1$ and multiplying the correction for a tauon in the loop
by 3 since this contributes the most of the 3 leptons. This leads
to $\rm a_{bh}<1.5 \times 10^{-10}$.\\
\section{Neutrino masses}
It is interesting that black hole process at the TeV scale can
induce small neutrino masses. Consider the process shown below:\\
\begin{figure}[h]
\centering
\includegraphics{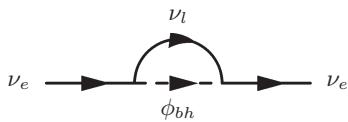}
\caption{Neutrino mass term} \label{fig:neut}
\end{figure}
\newline
This loop has a momentum integral:
\begin{equation}\rm
\int_0^{M_p} d^4k\frac{1}{k^2+M^2_{bh}}\frac{1}{k},
\end{equation}
which induces a mass $\rm m\sim g^2M_p$. Taking $\rm g$ of the
order of $\sim 10^{-6}$ would give masses of $\rm \sim 1~ eV$
size. If one is willing to believe that black hole processes at
the TeV range are the dominate source of neutrino mass then one
could place constraints on the $\rm g'_{ij}$ couplings using the
neutrino mass differences and
the mixing angles.\\
\section{Conclusion}
In this work we have shown that the phenomenology of a single
scalar black hole excitation at the TeV scale can be introduced
into the standard model without spoiling current experimental
results either for the muon magnetic moment or the neutrino
masses. All calculations made in this paper are done within the
effective non-renormalizable theory. One imagines that these
operators are remnant radiative effects of some high energy theory
like string theory. The true worth of our approach lies in its
ability to attain sensible results that are comparable with
experiment. Our approach has the flexibility to accommodate
alternative black hole phenomenology that would be dependent on
the results of forthcoming searches planned for the LHC.\\
\section*{Appendix 1}\label{Appendix:quantisation} In this appendix
we present the quantization scheme of a
charged spin zero black hole in n+3 spatial dimensions. \\
The full Kerr solution in higher dimensions is not analytically
tractable, nevertheless scalar excitations are the most important
in the present work and these we can solve for. In units with $\rm
\hbar=c=1$ we have the Einstein-Hilbert-Maxwell action:\\
\begin{equation}\rm
S=\frac{1}{16\pi G_{4+n}} \int
d^{n+4}x\sqrt{-g}\left(R-\mathcal{F_{\mu \nu}}\mathcal{F^{\mu
\nu}}\right).
\end{equation}
The solution \cite{Myers:1986un} is,\\
\begin{eqnarray}\label{eqn:RNF}\rm
f=g^{-1}
=(1-\frac{C}{r^{n+1}}+\frac{D^2}{r^{2(n+1)}})^{\frac{1}{2}},
\end{eqnarray}
where $\rm C=\frac{16\pi G_{4+n} M_{bh}}{S_{n+3}(n+2)}$ and
$\rm D^2=\frac{2Q^2G_{4+n}}{(n+2)(n+1)}$.\\
From equation (\ref{eqn:RNF}) there is an event horizon if:
\begin{eqnarray}\rm
r_{\pm}^{n+1}=\frac{1}{2}C \pm \sqrt{\frac{1}{4}C^2-D^2}.
\end{eqnarray} We now want to perform the canonical quantization
on the area of the outer horizon $\rm (A=A(r_{+}))$ in the same
spirit as Bekenstein \cite{Bekenstein:1974jk}. Recall that the
irreducible mass of a black hole, $\rm M_{ir}$, is related to its
area via:
\begin{eqnarray*}\rm
M_{ir}^2= \frac{A}{16\pi G_{4+n}^2}.
\end{eqnarray*}
Quantizing the irreducible mass $\rm M_{ir}=n_{b} g_{p}$ and the
charge $\rm Q=q e$ and rearranging for $\rm M_{bh}$ we find:
\begin{equation}\label{eqn:1}
\rm \frac{M_{bh}}{M_p}=c_{1}
(n_bg_{p})^{\frac{n+1}{n+2}}\left[1+\frac{1}{4}\frac{c_2 q^2
\alpha_{em} }{ (n_bg_{p})^{\frac{2n+2}{n+2}}}\right].
\end{equation}
Where,
\begin{eqnarray}\nonumber \rm
c_1&\equiv& (n+2)
\left(\frac{S_{n+3}}{16\pi}\right)^{\frac{1}{n+2}},\\\rm \nonumber
c_2 &\equiv & 2 \sqrt{2 (n+2)(n+1)}
\left(\frac{S_{n+3}}{16\pi}\right)^{\frac{2n+2}{n+2}},
\end{eqnarray}
$\rm S_m=\frac{2 \pi^{m/2}}{\Gamma(m/2)}$ is the surface area of a
unit m-sphere (ie in the case of an (n+3)-sphere $\rm
A=S_{n+3}r_{+}^{n+2}$) and $\rm n_b (\in \mathbbm{N})$ and q ($\in
\rm \mathbbm{Z}$) are the quantization numbers for mass and charge
respectively. $\rm n_{b}$ is not to be confused with n the number
of extra dimensions; we have adopted the notation used in
\cite{Bilke:2002rf}. The black hole mass gap, $\rm g_{p}$, is
controversial. Some authors use $\rm g_{p}=0.614/\pi$, which is
calculated in a loop quantum gravity
framework \cite{Khriplovich:2001je}.\\
In the n=0 case $\rm c_1$ and $\rm c_2$ are equal to one and we
recover the 3 dimensional quantization scheme:
\begin{equation}\nonumber
\rm \frac{M_{\rm
bh}}{M_{p}}=(n_bg_{p})^{1/2}\left[1+\frac{1}{4}\frac{q^2\alpha_{em}}{n_bg}\right].
\end{equation}

In quantizing, see equation (\ref{eqn:1}), we have introduced an
infinite tower of black holes labelled by the numbers $\rm n_{b}$,
q (and if we took the realistic case with angular momentum J also)
each with a definite mass, $\rm M_{bh}^{(n_{b},q,J)}$, and
therefore a different propagator. Thus, to calculate the $\cal{M}$
matrix for any given process we would need to sum over all the
black hole modes that can contribute. In the current work the
tower is naturally cut off at the $\rm M_p$ scale. The exact
number of modes that can participate will therefore depend on the
value of $\rm g_{p}$ and the number of
extra dimensions.\\
As $\rm \phi_{bh}$ is an effective 4 dimensional field for a
higher dimensional black hole Kaluza-Klein modes will also be
present of mass $\rm M_{bh}^n=\sqrt{M_{bh}^2+n^2/R}$ . For
simplicity we choose a black hole mass such
that the higher modes can be pushed above the cut off. \\
In this paper we have focused on the observable effects that a
single scalar excitation will produce. If in new experiments black
hole effects are discovered one can then use equation
(\ref{eqn:1}) to determine both $\rm g_{p}$ and $\rm n$ and
the related black hole phenomenology.\\

\acknowledgements JD wishes to thank K.L. McDonald for useful
discussions, SRC thanks the SERC, DST, India for support.
\bibliography{bhbib}

\end{document}